\documentclass[runningheads]{llncs}
\usepackage{amsmath, amssymb, gensymb, graphicx}
\usepackage{hyperref}
\begin{document}
\title{Automatic lesion segmentation and Pathological Myopia classification in fundus images\thanks{Supported by the Laboratory of Technological Innovation in Health of the Federal University of Rio Grande do Norte.}}
%
%
\author{Cefas Rodrigues Freire\inst{1} \and
Julio Cesar da Costa Moura\inst{1} \and
Daniele Montenegro da Silva Barros\inst{1} \and
Ricardo Alexsandro de Medeiros Valentim\inst{1}}
\authorrunning{C. Rodrigues et al.}
%
\institute{
Laboratory of Technological Innovation in Health, \\ 
Federal University of Rio Grande do Norte, Natal-RN, Brazil\\
\url{http://lais.huol.ufrn.br}}

\maketitle 

\begin{abstract}
In this paper we present algorithms to diagnosis Pathological Myopia (PM) and detection of retinal structures and lesions such as Optic Disc (OD), Fovea, Atrophy and Detachment. All these tasks were performed in fundus imaging from PM patients and they are requirements to participate in the Pathologic Myopia Challenge (PALM). The challenge was organized as a half day Challenge, a Satellite Event of The IEEE International Symposium on Biomedical Imaging in Venice Italy. Our method applies different Deep Learning techniques for each task. Transfer learning is applied in all tasks using Xception as the baseline model. Also, some key ideas of YOLO architecture are used in the Optic Disc segmentation algorithm pipeline. We have evaluated our model's performance according the challenge rules in terms of AUC-ROC, F1-Score, Mean Dice Score and Mean Euclidean Distance. For initial activities our method has shown satisfactory results.

\end{abstract}

\section{Introduction}\label{introduction}

In this work we present a Convolutional Neural Network (CNN) classification model to score Pathological Myopia (PM) risk in fundus imaging; an optic disc detection and segmentation algorithm using polar coordinates and a object detection model inspired on YOLO \cite{redmon2018yolov3} core concepts; fovea, atrophy and detachment detection method, both based in CNN models. All the tasks were performed to cover the four tasks in the PALM Challenge: 1) Classification of PM and non-PM; 2) Detection and segmentation of disc; 3) Localization of fovea and 4) Detection and Segmentation of retinal lesions. 

The remainder of this work is organized as follow:
Section \ref{datasets} describes all used databases to train and evaluate models. Section \ref{methods} presents the main concepts that are part of the proposed system, including the preprocessing steps, the model structure definition and the model training pipeline to each challenge task. The Section \ref{results} reports the results and discussion, as well the evaluation metrics. Finally, the Section \ref{conclusions} draw the conclusions and the future perspectives. 

\section{Datasets}\label{datasets}

In order to train and evaluate the proposed models, beyond the data provided by the challenge organization, the REFUGE \cite{REFUGE}, RIGA \cite{almazroa2018retinal} and IDRiD \cite{porwal2018indian} datasets were used. Below there is a short description of these datasets.

\subsection{REFUGE}

This dataset is part of the retinal fundus glaucoma challenge organized by the
Medical Image Computing and Computer Assisted Intervention (MICCAI) 2018 meeting. 
It is composed by 1,200 fundus images split 1:1:1 into a training set of 400 images, 
a validation set of 400 and a test set composed by 400 fundus. For each image in both training and validation sets, It is provided the glaucomatus corresponding label, as well the OD and OC annotations masks. The database can be found at \url{https://refuge.grand-challenge.org/}.

\subsection{RIGA}

The RIGA stands for Retinal Images for Glaucoma Analysis. It is a retinal fundus images dataset of glaucoma patients. For each image, It is provided contour annotation of Optic Disc and Optic Disc Cup segmented by 7 different ophtalmologists. The dataset was developed through MESSIDOR images concatenation and images from from Magrabi Eye Centre, Riyadh and Bin Rashid Ophthalmic Centre, Riyadh. This database is released on Creative Commons 4.0 and is available at \url{https://deepblue.lib.umich.edu/data/concern/data\_sets/3b591905z}.

\subsection{IDRiD}

The  IDRiD dataset consists of 516 fundus images and ground truths associated with the signs of Diabetic Retinopathy (DR) and Diabetic Macular Edema(DME). This database provides: pixel level annotations of typical DR lesions and OD; image level disease severity grading of DR, and DME; OD and Fovea center coordinates. These categories are split into training and validation sets and named as ”Segmentation”, ”Disease Grading” and ”Localization”, respectively.

\section{Methods}\label{methods}

\subsection{Classification of PM and non-PM}\label{sec:classificacao_pm}

To perform this task, it was used the Transfer Learning technique due its feasibility related to the provided data and necessary time.

\subsubsection{Model architecture}\label{sec:classificacao_pm_ma}
The Xception \cite{chollet2017xception} was chosen as the baseline architecture due to the number of parameters and its performance in ImageNET. It was used the Xception ImageNET pretrained weights and all layers until the last Separable Convolutional Layer. On its top was appended a Global Average Pooling layer, followed by a fully-connected layer with a single output activated by the sigmoid function, corresponding to the model's guess of PM classification.

\subsubsection{Data preparation}\label{sec:classificacao_pm_dp}

It was used 749 images from the RIGA database, 800 images of the training and validation set of REFUGE database and the complete training set of provided data. In order to prepare data to model training to each image was assigned a PM or non-PM label. To feed the data to the model training each image was assigned as PM or Non-PM. Once that the RIGA and REFUGE do not provide labels for PM, all the images from both databases were labeled as non-PM. For the provided data, all High Myopia (HM) and normal images were defined as non-PM and the last ones as PM. Next, It was applied a normalization step which ensures all images of size 299$\times$299 pixels, cropping a rectangular centered image before resize if necessary. Finally all the pixels values were linearly rescaled between -1 and 1.

\subsubsection{Loss function}\label{sec:classificacao_pm_lf}

It was used the binary cross entropy loss function. Considering that each image (I) assigned label (non-PM or PM) is encoded to $y^{(i)} \in \{0, 1\}$ and the model's output PM probability related to I is ${\hat{y}}^{(i)} \in (0, 1)$, then the loss function to $m$ images is defined by the following formula:
\begin{equation}
L_{class}(y,\hat{y})=-\frac{1}{m}\sum_{i=1}^m y^{(i)}\log({\hat{y}}^{(i)})+(1-y^{(i)})\log(1-{\hat{y}}^{(i)})
\end{equation}
where $y=\{y^{(1)}\dots y^{(m)}\}$ and $\hat{y}=\{{\hat{y}}^{(1)}\dots {\hat{y}}^{(m)}\}$.
\subsubsection{Training}
Before to train, the processed images and targets are split on training and validation sets according the 8:2 proportion (20\% of images to the validation set and 80\% to the training set). This procedure is performed in such way that the proportion of non-PM and PM images in each set was the same. The total number of training and validation images are respectively 1190 and 299.

At training time It was used the Adamax \cite{kingma2014adam} optimizer, with $\beta_1=0.9$, and $\beta_2=0.999$. The learning rate is $0.05$ and without any decay. The batch size used was 20 and the momentum of batch normalization was $0.9$. No data augmentation was performed and the corresponding weights of first pretrained 50 Xception layers are not updated by the gradients. In the other layers It was applied a L2 regularization factor of $0.02$. The training was performed throughout 4 epochs. 

\subsection{Optic Disc Segmentation}\label{sec:segmentacao_od}

To perform this task the algorithm pipeline was divided into four steps: detect OD center location in the whole image, extract a squared ROI in the found location, detect the OD contour boundary in the ROI and generate the annotation mask from the contour. So, two models were used: a model to perform the localization and other to automatic find the OD boundary.

\subsubsection{Data Preparation}\label{sec:segmentacao_od_dp}
It was used used 81 images from the segmentation set in the IDRiD database; All the 400 images from provided training set; All the the 800 training images from the REFUGE and 749 images from the RIGA. The prepossessing consisted in images scaling to $229\times 229$ pixels. If any image is rectangular shape, a centered squared cropping was performed before the imaging rescaling. The $(c_x, c_y)$ optic disc center was found using the annotation provided by the databases. Then the ROI is extracted cropping a region of 224x224 pixels centered in the OD center of a equivalent rescaled $800\times 800$ pixels image. From  the data annotation available on the datasets, the contour points related to the OD were extracted to each image. For the REFUGE dataset only the annotation provided by  the  medical  expert  was  used.

In order to acquire data for the boundary detection model, the $(x,y)$ contour points extracted from each ROI $(j)$ were used to compute a function which represents the contour as a polar function.  It is mathematically defined as follows: 
\begin{equation}\label{eq:contornons}
Y_i^{(j)} = \sum_{k=1}^{(N+1)^2}{\beta^{(j)}}_k{\cos(\theta_{i}^{(j)})}^{c_k}{\sin(\theta_{i}^{(j)})}^{s_k}
\end{equation}
in which $Y_i^{(j)}$ is the radius related of the $\theta_i^{(j)}$ angle; Each $\theta_i^{(j)}$ angle was extracted by the arc tangent the $(x,y)$ Cartesian coordinates of every single contour point of the training example $(j)$. $c_i$ and $s_i$ are all the possible permutations between any two numbers chosen among $0$ and $N=5$. 

In this context, $Y_i^{(j)}$ is described as:

\begin{equation}\label{eq:y_contornos}
Y^{(j)} = X^{(j)}\beta^{(j)} + \epsilon
\end{equation}
in which $X^{(j)}$ is:
\begin{equation}
X^{(j)}=\begin{bmatrix}
\cos(\theta_{1}^{(j)})^{c_1}\sin(\theta_{1}^{(j)})^{s_1}&\cdots &\cos(\theta_{1}^{(j)})^{c_n}\sin(\theta_{1}^{(j)})^{s_n} \\
\cos(\theta_{2}^{(j)})^{c_1}\sin(\theta_{2}^{(j)})^{s_1}&\cdots &\cos(\theta_{2}^{(j)})^{c_n}\sin(\theta_{2}^{(j)})^{s_n} \\
\vdots & \vdots & \vdots\\
\cos(\theta_{m}^{(j)})^{c_1}\sin(\theta_{m}^{(j)})^{s_1}&\cdots
&\cos(\theta_{m}^{(j)})^{c_n}\sin(\theta_{m}^{(j)})^{s_n}
\end{bmatrix}
\end{equation}

So, the parameters $\hat{\beta}^{(j)}$ are obtained by the normal equation solution, as follows:
\begin{equation}
\hat{\beta}^{(j)} = ({X^{(j)}}^{T}X^{(j)})^{-1}{X^{(j)}}^{T}Y^{(j)}
\end{equation}
in which $\hat{\beta}^{(j)}$ symbol means the $\beta^{(j)}$ estimation.

\subsubsection{Model Architecture}
The localization model was based on the YOLO and Xception architectures. The Xception model was used in the same way as the PM and non-PM classification, however on the top, it was added 3 $1\times 1$ convolutional layers with sigmoid activation function and $depth=1$. The concatenated outputs of the convolutional layers were based on the YOLO method, which for each grid cell there is a probability of the cell to possess the object center, and x-y displacements based on the top-left corner of that cell. So, the final model output shape to a given example is $10\times 10\times 3$. The highest cell's probability has considered to get the $({c^*}_x,{c^*}_y)$ points of predicted OD center. 

The Table \ref{tab:tabela1} shows the boundary detection model architecture. 
\begin{table}[h]
\caption{Model Structure}
\label{tab:tabela1}
\begin{center}
 \begin{tabular}{|c c c c |}
 \hline
 & Type & Filters & Size \\ [1 ex] 
 \hline
 \hline
 & Convolucional & $3 \times 3$ & 16  \\ [1 ex]
 \hline
 & MaxPooling+Dropout & $2 \times 2$ &   \\ [1 ex]
 \hline
 \hline
 & Convolucional & $ 3 \times 3 $ & 32  \\ [1 ex]
 \hline
 & MaxPooling+Dropout & $2 \times 2 $ &   \\ [1 ex]
\hline
\hline
 $2 \times $ & Convolutional+Dropout & $3 \times 3$ & 32 \\ [1 ex]
 \hline
 \hline
 & Convolutional & $3 \times 3 $ & 32  \\ [1 ex]
 \hline
 & MaxPooling+Dropout & $ 2 \times 2 $ &   \\ [1 ex]
 \hline
 \hline
 $2 \times $ & Convolutional+Dropout & $3 \times 3 $ & 64 \\ [1 ex]
  \hline
  \hline
 & Convolutional & $3 \times 3 $ & 64  \\ [1 ex]
 \hline
 & MaxPooling+Dropout & $ 2 \times 2 $  &   \\ [1 ex]
 \hline
 \hline
 & Convolutional+Dropout & $ 3 \times 3 $& 128 \\ [1 ex]
 \hline
 \hline
 & Convolutional & $3 \times3 $& 128  \\ [1 ex]
 \hline
 & MaxPooling+Dropout & $2 \times 2 $ &   \\ [1 ex]
 \hline
 \hline
 $2 \times $ & Convolutional & $1 \times 1 $& 64 \\ [1 ex] 
 \hline \hline
 & Convolutional & 1x1 & 32 \\ [1 ex] 
 \hline \hline
 $4 \times $ & Fully Connected+Dropout &  & 500 \\ [1 ex] 
 \hline \hline
 & Fully Connected+Dropout &  & 1000 \\ [1 ex] 
 \hline
 & Fully Connected &  & 38 \\ [1 ex] 
 \hline
\end{tabular}
\end{center}
\end{table}

\subsubsection{Loss Function}
In order to enable the localization model's loss computation, it was generated ground truth values for the wished model output for each image. The general loss function for the localization was defined as weighted sum of the individual loss function for classification, and x-y displacements. To the x-y displacements the mean squared Euclidean distance was used as loss function for all the cases which there is the optic disc center in cell which the x-y displacement is related. For the classification it was used a modified form of the cross entropy according with some ideas in \cite{lin2017focal}. It is necessary to deal with the class balance.

Considering that each $(j, k)$ image grid cell OD detection probability is encoded to ${c^i_{j,k}} \in \{0, 1\}$ and the model's output probability related to that cell is is ${\hat{c}_{j,k}}^i \in (0, 1)$, then the classification loss function to $m$ images is defined by the following formula:
\begin{equation}
\begin{split}L_{c}(c,\hat{c})=-\frac{1}{m}\sum_{i=1}^m \sum_{j=1}^{10} \sum_{k=1}^{10} 0.75c^i_{j,k}\log({\hat{c}_{j,k}}^i)+
0.25(1-{c^i_{j,k}})\log(1-{\hat{c}_{j,k}}^i)
\end{split}
\end{equation}
where $c$ is the set of all ${c^i_{j,k}}$ values and $\hat{c}$ is the set of all ${\hat{c}_{j,k}}^i$ values with $i\in\{1, 2, \dots, m\}$, $j\in\{1, 2, \dots, 10\}$ and $k\in\{1, 2, \dots, 10\}$.

Adopting the same convention as before, considering that each $(j, k)$ image grid cell x-y displacements are encoded to ${x^i_{j,k}} \in (0, 1)$ and ${y^i_{j,k}} \in (0, 1)$ and the model's output displacements related to that cell is is ${\hat{x}^i_{j,k}} \in \{0, 1\}$ and ${\hat{y}^i_{j,k}} \in \{0, 1\}$, then the displacement loss function to $m$ images is defined by the following formula:
\begin{equation}
\begin{split}
L_{d}(x,y,\hat{x},\hat{y})=-\frac{1}{m}\sum_{i=1}^m \sum_{j=1}^{10} \sum_{k=1}^{10} [({x^i_{j,k}} - {\hat{x}^i_{j,k}})^2 + ({y^i_{j,k}} - {\hat{y}^i_{j,k}})^2]{{y}^i_{j,k}}
\end{split}
\end{equation}

The overall localization model's loss function for $m$ trained example is defined by: 
\begin{equation}
\begin{split}
L(c,x,y,\hat{c},\hat{x},\hat{y}) = 2 L_{c}+L_{d}
\end{split}
\end{equation}

Before to compute the loss function used in the boundary detection model training, a set of 72 angles (from 0\degree to 360\degree) equally split were generated. For each training example $(j)$ these angles were applied in Equation \ref{eq:contornons} to compute the $Y_i^{(j)}$ radius associated to that example. These angle-radius pairs were converted in Cartesian coordinates and summed with the original angle center $({c_x}^{(j)},{c_y}^{(j)})$ coordinates. The same computation was applied to the model's outputs using the first 36 values $\hat{\beta^{*}}^{(j)}$ to calculate the radius and the last 2 as centers $({{c^*}_x}^{(j)},{{c^*}_x}^{(j)})$ to calculate Cartesian coordinates. The summation of the square of the Euclidean distance between the Cartesian ground truth and the Cartesian predicted is so defined:

\begin{equation}
L(\hat{Y}_i^{(j)}, \hat{Y^*}_i^{(j)})=f_1(\hat{Y}_i^{(j)}, \hat{Y^*}_i^{(j)})+f_2(\hat{Y}_i^{(j)}, \hat{Y^*}_i^{(j)})
\end{equation}
in which $\hat{Y^*}_i^{(j)}$ is the approximation of $\hat{Y}_i^{(j)}$ computed by the model output $\hat{\beta^{*}}^{(j)}$ and:

\begin{equation}
f_1(\hat{Y}_i^{(j)}, \hat{Y^*}_i^{(j)})=\sum_{k=1}^{72}[\cos(\Theta_{k}\hat{\beta}^{(j)})+{c_x}^{(j)}-\cos(\Theta_{k}\hat{\beta^{*}}^{(j)})-{{c^*}_x}^{(j)}]^2
\end{equation}
and
\begin{equation}
f_2(\hat{Y}_i^{(j)}, \hat{Y^*}_i^{(j)})=\sum_{k=1}^{72}[\sin(\Theta_{k}\hat{\beta}^{(j)})+{c_y}^{(j)}-\sin(\Theta_{k}\hat{\beta^{*}}^{(j)})-{{c^*}_y}^{(j)}]^2
\end{equation}

The $\Theta$ is a $72\times (N+1)^2$ matrix in which each row $\Theta_{k}$ is defined by:

\begin{equation}
\Theta_{k}=\begin{bmatrix}
\cos(\theta_{k})^{c_1}\sin(\theta_{k})^{s_1}\\ 
\vdots \\
\cos(\theta_{k})^{c_n}\sin(\theta_{k})^{s_n}
\end{bmatrix}^{T}
\end{equation}

The overall cost function used is then defined by equation:
\begin{equation}
L_{total}(\hat{Y}_i, {\hat{Y}^*}_i)=\frac{1}{m}\sum_{j=1}^{m} L(\hat{Y}_i^{(j)},\hat{Y^*}_i^{(j)})
\end{equation}

\subsubsection{Training}
On the localization model training step, 80\% of the images were used to train and 20\% to validation. The model was trained for 20 epochs, with learning rate of $0.005$. The AdamMax optimization algorithm was used with $\beta_1$ of $0.9$ and $\beta_2$ of $0.999$.  The batch size used was 20 and the momentum of batch normalization was $0.9$. horizontal flips augmentation was performed and the corresponding weights of first pretrained 50 Xception layers are not updated by the gradients. In the other layers It was applied a L1 and L2 regularizations with factor of $0.1$ for L1 and $0.25$ for L2. The training was performed throughout 20 epochs.

On the boundary detection model training step, 85\% of the images were used to train and 15\% to validation. The model was trained for 200 epochs, with learning rate of $0.0001$.  Horizontal and vertical displacements augmentation were performed. It was applied and L2 regularization with factor of $0.05$ and 20\% of input units drop was applied to all dropout layers. The training was performed throughout 200 epochs.

It was chosen the best model performance according the best performance in the validation set.

\subsection{Localization of fovea}\label{sec:localizacao_fovea}
\subsubsection{Data Preparation}
It was used almost the same method of the Section \ref{sec:segmentacao_od_dp}, the only difference is that it was used the localization set from the IDRiD (413 images), REFUGE (800 images from the training and validation set) and all images from the challenge dataset.

\subsubsection{Model Architecture}
It was used almost the same architecture defined in the Section \ref{sec:classificacao_pm_ma}. The only difference is that 2 model outputs were computed with sigmoid activation function whose the values were multiplied by 299 providing the model output. These output values are related to model's guess of (x,y) coordinate of the fovea in the $299\times 299$ pixels image.

\subsubsection{Loss Function}
It was used as loss function the mean of the euclidean between model's output and the equivalent ground truth (fovea center computed by $m$ provided images annotations). The formula is defined by:

\begin{equation}
    L=\frac{1}{m}\sum_{i=1}^m\sqrt{(x^i-\hat{x}^i) - (y^i-\hat{y}^i)}
\end{equation}
in which $(x^i, y^i)$ are the ground truth location of fovea related to the training example $i$ and the $(\hat{x}^i, \hat{y}^i)$ are the model's guess associated to that example.

\subsubsection{Training}
The training step was almost the same of the optic disc segmentation, but with 25 epochs.

\subsection{Detection and Segmentation of Atrophy}\label{sec:segmentacao_atrofia}
\subsubsection{Data Preparation}
For this task it was used only the training set images provided by the challenge organization. 

The images were normalized to $512\times 512$ pixels, using the same square cropping method which was used in the segmentation of the optic disc. The pixel level normalization was also the same of the performed in the section \ref{sec:classificacao_pm_dp}.

\subsubsection{Model Architecture}
It was used a adaptation from the DeepLabV3+ \cite{chen2018encoder}, adapted to 2 classes. The first class is related to the presence of Atrophy and the last one is the background. It was used a SoftMax ativation.

\subsubsection{Loss Function}
The categorical crossentropy was used as loss function. 

\subsubsection{Training}
On the training step, 80\% of the images were used to train and 20\% to validation. The model was trained for 40 epochs, with learning rate of 0.005. The AdamMax was used with $\beta 1$ of 0.9 and $\beta 2$ of 0.999.  The batch size used was 2 and the momentum of batch normalization was $0.9$. rotation, horizontal and vertical scaling and flip horizontal augmentation was performed and the corresponding weights of first pretrained 50 Xception layers are not updated by the gradients. The training was performed throughout 20 epochs.

It was chosen the best model performance according the validation set.

\subsubsection{Masks Determination}
To improve the results the original image is fed to model firstly and the result is stored, secondly this image is horizontally flipped and fed to the model. Thirdly, the respective result is horizontally flipped to be equivalent to the result from the original image. Next it is computed a mean between both results. Then the argmax function is applied to the mean, which returns only 0 or 1 values, related to atrophy and background, respectively. This values is used to create the mask according to the challenge rules, however the masks dimensions are still $512\times 512$. To back this results to the original shape, it is applied a reverse normalization.

\subsection{Detection and Segmentation of Detachment}\label{sec:segmentacao_descolamento}
In this step we did not try solve the segmentation problem itself, due the small number of images with detachment. Our model results a probability of the image have or not the detachment. The respective mask, in case of detection is defined as the lesion is present in all image. 

As a mask, It was used a center positioned circumference in the image center whose diameter corresponds to the image width. All the pixels inside the circumference. are attributed with zero values. Outside this circumference the pixels were attributed with 255 values.

\subsubsection{Data Preparation}
It was performed the same method that was used in the section \ref{sec:classificacao_pm_dp} only for the database available from the organization.

\subsubsection{Model Architecture}
It was used the same method which was used in the Section \ref{sec:classificacao_pm_ma}.

\subsubsection{Loss Function}
The categorical crossentropy was used as loss function. 

\subsubsection{Training}

On the training step, 80\% of the images were used to train and 20\% to validation. The model was trained for 40 epochs, with learning rate of 0.005. The AdamMax was used with $\beta 1$ of 0.9 and $\beta 2$ of 0.999.  The batch size used was 2 and the momentum of batch normalization was $0.9$. rotation, horizontal and vertical scaling and flip horizontal augmentation was performed and the corresponding weights of first pretrained 50 Xception layers are not updated by the gradients. The training was performed throughout 20 epochs. It was applied a L1 normalization with factor of 0.1.

\section{Results and Discussion}\label{results}
To evaluate the proposed method we used the metrics suggested in the challenge rules: Area Under ROC curve (AUC-ROC) to evaluate the \ref{sec:classificacao_pm}; F1-Score (F1) and Mean Dice (MD) coefficient were used to evaluate the \ref{sec:segmentacao_od}, \ref{sec:segmentacao_atrofia} and \ref{sec:segmentacao_descolamento}. Finally, for the \ref{sec:localizacao_fovea} it was used the mean Euclidean Distance (ED). It  was also performed a comparison between these metrics computed on the test dataset and the train dataset. These results are shown on \ref{tab:tabela2} and \ref{tab:tabela3}

The Table \ref{tab:tabela2} shows the evaluation results. 
\begin{table}[h]
\caption{Evaluation Results}
\label{tab:tabela2}
    \begin{center}
        \begin{tabular}{| c c c c c c|} \hline
	    & & \textbf{AUC-ROC} & \textbf{F1-Score} & \textbf{MD-Score} & \textbf{Mean ED} \\ \hline
	    \textbf{PM and non-PM} & & 0.9984 & ----- & ----- & ----- \\ \hline
	    \textbf{Optic disc segmentation} & & ----- & 0.9934 & 0.9329 & ----- \\ \hline
	    \textbf{Fovea Detection} & & ----- & ----- & ----- & 79.2701 \\ \hline
	    \textbf{Atrophy Segmentation} & & ----- & 0.8842 & 0.7942 & ----- \\ \hline
	    \textbf{Detachment Segmentation} & & ----- & 0.7121 & 1.0000 & ----- \\ \hline
	    
        \end{tabular}
      \end{center}
\end{table}

The Table \ref{tab:tabela3} shows the test results. 
\begin{table}[h]
\caption{Test Results}
\label{tab:tabela3}
    \begin{center}
        \begin{tabular}{| c c c c c c|} \hline
	    & & \textbf{AUC-ROC} & \textbf{F1-Score} & \textbf{MD-Score} & \textbf{Mean ED} \\ \hline
	    \textbf{PM and non-PM} & & 0.9957 & ----- & ----- & ----- \\ \hline
	    \textbf{Optic disc segmentation} & & ----- & 0.9855 & 0.9092 & ----- \\ \hline
	    \textbf{Fovea Detection} & & ----- & ----- & ----- & 71.2958 \\ \hline
	    \textbf{Atrophy Segmentation} & & ----- & 0.9091 & 0.7798 & ----- \\ \hline
	    \textbf{Detachment Segmentation} & & ----- & 0.7273 & 0.5547 & ----- \\ \hline
	    
        \end{tabular}
      \end{center}
\end{table}


\section{Conclusions}\label{conclusions}
In this paper we have proposed a set of methods to classify Pathological Myopia as well Atrophy and Detachment. The algorithms developed also identify ocular structures as: Optic Disc and Fovea. 
The next steps consist to improve the results and acquire new data.

\bibliographystyle{splncs04}
\bibliography{references}

\end{document}